\documentclass[twocolumn,a4paper,aps,prb,draft]{revtex4}
\usepackage{amsmath}
\usepackage[final]{graphics}

\begin{document}

\title{Comment on ``Spherical 2 + \textit{p} spin-glass model: An
analytically solvable model with a glass-to-glass transition'' }

\author{V. Krakoviack}

\affiliation{Laboratoire de Chimie, UMR CNRS 5182, {\'E}cole Normale
Sup{\'e}rieure de Lyon, 46 All{\'e}e d'Italie, 69364 Lyon Cedex 07,
France}

\date{\today}

\begin{abstract}
Guided by old results on simple mode-coupling models displaying
glass-glass transitions, we demonstrate, through a crude analysis of
the solution with one step of replica symmetry breaking (1RSB) derived
by Crisanti and Leuzzi for the spherical $s+p$ mean-field spin glass
[Phys. Rev. B \textbf{73}, 014412 (2006)], that the phase behavior of
these systems is not yet fully understood when $s$ and $p$ are well
separated. First, there seems to be a possibility of glass-glass
transition scenarios in these systems. Second, we find clear
indications that the 1RSB solution cannot be correct in the full
glassy phase. Therefore, while the proposed analysis is clearly naive
and probably inexact, it definitely calls for a reassessment of the
physics of these systems, with the promise of potentially interesting
new developments in the theory of disordered and complex systems.
\end{abstract}

\maketitle

In a recent paper \cite{CriLeu06PRB}, Crisanti and Leuzzi have
proposed a detailed analysis of the so-called spherical $s+p$
mean-field spin-glass model, defined by the Hamiltonian
\begin{equation*}
\mathcal{H} = \sum_{i_1<\cdots <i_s}^{1,N} J^{(s)}_{i_1\ldots i_s}
\sigma_{i_1} \cdots \sigma_{i_s} + \sum_{i_1<\cdots<i_p}^{1,N}
J^{(p)}_{i_1\ldots i_p} \sigma_{i_1} \cdots \sigma_{i_p},
\end{equation*}
where $s$ and $p$ are integers such that $2 \leq s<p$, the spins
$\sigma_i$ are $N$ real variables subject to the spherical constraint
$\sum_{i=1}^{N} \sigma_i^2 = N$, and the random coupling constants
$J^{(s)}_{i_1\ldots i_s}$ and $J^{(p)}_{i_1\ldots i_p}$ are
uncorrelated zero mean Gaussian variables with variances
$s!J_s^2/2N^{s-1}$ and $p!J_p^2/2N^{p-1}$, respectively. At the
inverse temperature $\beta$, these variances enter into the definition
of the control parameters of the model, $\mu_s= (\beta J_s)^2 s/2$ and
$\mu_p= (\beta J_p)^2 p/2$.

In fact, Ref.~\onlinecite{CriLeu06PRB} mostly deals with the case
$s=2$, $p\geq4$, which displays a rich phase diagram with four
different phases and a variety of transitions between them
\cite{CriLeu04PRL}. The case $2<s<p$, to which the present Comment is
devoted, is only discussed shortly in an appendix and illustrated with
the $3+4$ model, whose static and dynamical phase diagrams both
exhibit only two phases, paramagnetic with replica symmetry (RS) and
glassy with one step of replica symmetry breaking (1RSB), separated by
smooth transition lines. This result is claimed to be generic.

In this Comment, we show that this might not be the case and that
there could be a parameter domain apparently left unexplored by all
previous workers and corresponding to $2<s<p$ and $p-s$ large enough,
where new phenomena occur. To achieve this goal, we first recall
seemingly little known results about simple schematic models of the
mode-coupling theory (MCT) for the liquid-glass transition
\cite{LesHouches,GotSjo92RPP,Got99JPCM}, which show glass-glass (G-G)
transitions and higher-order singularities. Then, guided by the
insight gained within the framework of the MCT, we report the results
of a naive investigation of the 1RSB solution derived by Crisanti and
Leuzzi, which raise a number of issues about the structure of the
phase diagram of the systems with large $p-s$.

In the framework of the MCT \cite{LesHouches,GotSjo92RPP,Got99JPCM},
the schematic models are minimal models reproducing the typical
nonlinearities and bifurcation scenarios of the dynamical equations
generically derived within this theory. A particularly important class
consists of the so-called $F_{pq}$ models, $1\le p<q$, defined by the
memory kernels $m(t)=v_p \phi(t)^p + v_q \phi(t)^q$. They have been
widely studied and, although most of the works have dealt with the
$F_{12}$ and $F_{13}$ models\cite{LesHouches}, scattered results are
also available on more general $F_{pq}$ models
\cite{FucGotHofLat91JPCM,KraAlb02JCP}.

\begin{figure}
\includegraphics*{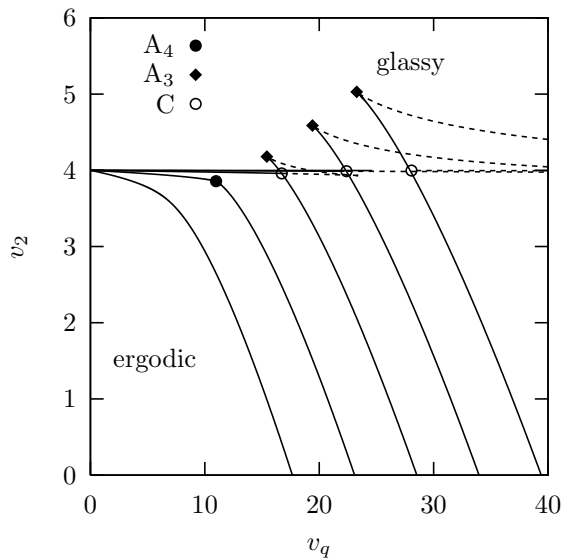}
\caption{\label{figmct} Phase diagrams of the $F_{pq}$ schematic
  models with $p=2$ and $q=7$, $9$, $11$, $13$, and $15$ (from left to
  right). The solid lines are the true transition lines and the dashed
  lines are the portions of the $\{v_p(f),v_q(f)\}$ parametric curves
  which are excluded by the dynamical stability criterion. The
  higher-order singularities (A$_3$ and A$_4$) and the crossing points
  (C) delimiting the G-G transition lines are indicated by symbols.}
\end{figure}

The bifurcation analysis of the solutions of the $F_{pq}$ equation
with $1<p<q$ leads to a parametric representation of the dynamical
transition line, which uses the non-ergodicity parameter
$f=\lim_{t\to+\infty} \phi(t)$ as the variable. It reads
\begin{equation*}
v_p(f)=\frac{1}{q-p}\frac{q-1-qf}{f^{p-1}(1-f)^2},
v_q(f)=\frac{1}{p-q}\frac{p-1-pf}{f^{q-1}(1-f)^2},
\end{equation*}
with $(p-1)/p \le f \le (q-1)/q$.  Topological changes occur in this
curve depending on $p$ and $q$, or more precisely on their combination
\cite{lambda}
\begin{equation*}
\Lambda(p,q)=\frac{\sqrt{pq} - \sqrt{(p-1)(q-1)}}{\sqrt{2}}.
\end{equation*}
If $p$ and $q$ are close enough, such that $\Lambda(p,q) < 1$, the
curve $\{v_p(f),v_q(f)\}$ is smooth, but, if they are widely
separated, such that $\Lambda(p,q) > 1$, a loop appears in the glassy
domain. Using then a dynamical stability criterion which states that,
when several values of the non-ergodicity parameter seem possible at a
given point $(v_p,v_q)$, only the largest one is physically meaningful
\cite{LesHouches}, one shows that two of the three branches of the
loop have to be discarded, while the remaining one forms a G-G
transition line which terminates at an A$_3$ higher-order singularity
(the ordinary MCT bifurcation is of type A$_2$) \cite{LesHouches}. In
the marginal cases like the models with $p=2$ and $q=9$ or $p=9$ and
$q=50$, where $\Lambda(p,q)=1$, the loop reduces to a point and yields
an A$_4$ singularity.  These results are illustrated in
Fig.~\ref{figmct} for different models with $p=2$.  Interestingly,
exactly the same dynamical scenarios are obtained in MCT studies of
colloidal suspensions with short-ranged attractions
\cite{DawFofFucGotSciSpeTarVoiZac01PRE,colloids}.

Let's now come back to the original spin-glass problem. One of the
motivations for the study of the mean-field spherical spin-glass
models with multispin interactions has been the finding that the time
evolution of the spin correlation function in the Langevin dynamics of
these systems
\cite{KirThi87PRL,CriHorSom93ZPB,BouCugKurMez96PA,CiuCri00EL} is
described at high temperature by schematic mode-coupling equations.
Namely, the high-temperature dynamics of the $s+p$ spin glass is
described by the $F_{(s-1)(p-1)}$ schematic model, with the
replacement $v_{s-1}=\mu_s$ and $v_{p-1}=\mu_p$.  This analogy turned
out to be a fruitful source of new theoretical developments in the
physics of structural glasses, allowing to complement the MCT,
originally a purely dynamical theory derived under the assumption of
equilibrium, with very interesting static \cite{CriSom92ZPB} and aging
\cite{CugKur93PRL} scenarios.  So, since the $F_{pq}$ schematic models
with well separated indices show interesting and nontrivial
bifurcation scenarios in their glassy domain, one might legitimately
wonder whether similar features could be present in the glassy phase
of the $s+p$ spin glass as studied by Crisanti and Leuzzi
\cite{CriLeu06PRB}.

We first consider the statics. Its method of solution within the 1RSB
scheme is described in detail in Ref.~\onlinecite{CriLeu06PRB} and we
only recall the expression of the free energy per spin $\Phi$ in the
1RSB phase as a function of the overlap parameter $q$ and of the usual
1RSB parameter $m$.  It reads $-\beta \Phi(q,m) = s_{\infty} +
G(q,m)$, where $s_\infty$ is the entropy per spin at infinite
temperature and $G(q,m)$ is given by
\begin{multline*}
2 G(q,m) = \frac{\mu_s}{s} + \frac{\mu_p}{p} + (m-1)
\left(\frac{\mu_s}{s} q^s + \frac{\mu_p}{p} q^p \right) + \\
\frac{1}{m} \ln (1-q+mq) +\frac{m-1}{m}\ln (1-q).
\end{multline*}
In this expression, the coupling constants $\mu_s$ and $\mu_p$ are
functions of $q$ and $m$ as well, obtained from the stationarity
conditions $\partial_q G(q,m) = 0$ and $\partial_m G(q,m)=0$.

\begin{figure}
\includegraphics*{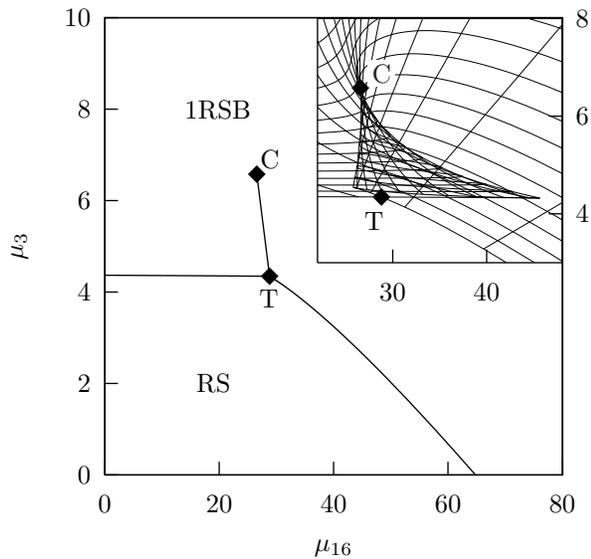}
\caption{\label{figstat} Putative 1RSB static phase diagram of the
  $3+16$ model in the $(\mu_3,\mu_{16})$ plane. T and C denote the
  ``triple'' and ``critical'' points and delimit the candidate G-G
  transition line. Insert: Projection onto the $(\mu_3,\mu_{16})$
  plane of the parametric surface $\{\mu_3(q,m), \mu_{16}(q,m),
  G(q,m)\}$ in the region of the candidate G-G transition line. The
  outline of a swallowtail is clearly visible.}
\end{figure}

In the same spirit as in the schematic MCT calculation, one is led to
consider the parametric surface $\{\mu_s(q,m),\mu_p(q,m),G(q,m)\}$,
which provides one with an implicit representation of the free energy
as a function of the control parameters, and to look for topological
changes when $s$ and $p$ are varied. As illustrated for the $3+16$
model in the insert of Fig.~\ref{figstat}, where a part of this
surface is plotted, such changes indeed occur when $p-s$ is increased.
A swallowtail, where the free energy as a function of $\mu_3$ and
$\mu_{16}$ would be multivalued, is clearly visible above the
surface. This looks very much like the loops met in the van der Waals
theory of phase coexistence. Following this analogy, since the
saddle-point calculation of the free energy requires that $G$ should
be minimized for given $\mu_s$ and $\mu_p$, a crude way of dealing
with this feature of the surface would be to discard it in order to
define uniquely the free energy at a given state point. Then, this
leaves one with a line of double points (a ``binodal''), along which
two distinct $(q,m)$ pairs correspond to the same value of the free
energy of the glassy phase. When this line is crossed, by following a
path of constant $\mu_3$ for instance, a discontinuity in $(q,m)$
occurs, while the free energy remains continuous. At the level of the
present discussion, this line appears as a candidate for a line of
static G-G transitions, starting at a ``triple'' point, where the
paramagnetic phase and two 1RSB phases with $m=1$ and different values
of $q$ would have the same free energy, and ending at a ``critical''
point, where the doublet $(q,m)$ is such that $\partial_q \mu_s
\partial_m \mu_p - \partial_m \mu_s \partial_q \mu_p = 0$ and
$\partial_q (\partial_q \mu_s \partial_m \mu_p - \partial_m \mu_s
\partial_q \mu_p) = 0$ (the first equality characterizes ``spinodals''
corresponding to the edges of the swallowtail and meeting at the
``critical'' point).

The resulting static phase diagram for the $3+16$ model would then be
the one outlined in Fig.~\ref{figstat}.  Its shape seems generic for
well separated $s$ and $p$, for instance, $s=3$ and $p\geq 13$, $s=4$
and $p\geq 23$, or $s=10$ and $p\geq 105$. The larger $s$, the larger
$p-s$ has to be for a candidate static G-G transition line to exist.

We now turn to the 1RSB solution for the dynamics of the $s+p$ model.
Following Crisanti and Leuzzi, the discussion is not based on a direct
analysis of the off-equilibrium dynamics of the system, but on the
study of its complexity $\Sigma$, a procedure which should yield
identical results. Thus it amounts here again to the consideration of
a set of parametric equations, where $\mu_s$, $\mu_p$, and $\Sigma$
are expressed as functions of the dynamical overlap and 1RSB
parameters. See Ref.~\onlinecite{CriLeu06PRB} for the corresponding
expressions.

\begin{figure}
\includegraphics*{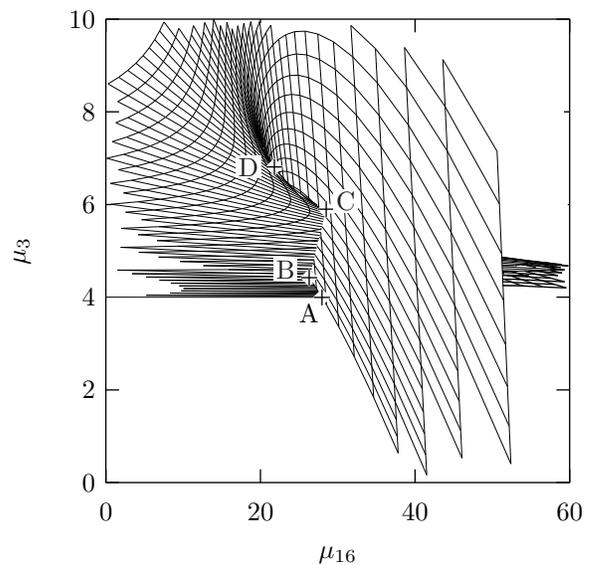}
\caption{\label{figdyn} Top view of the parametric surface
  $\{\mu_3(q,m), \mu_{16}(q,m), \Sigma(q,m)\}$ for the dynamical 1RSB
  phase of the $3+16$ model. Except in the rightmost part of the
  figure where the surface is cut off in order to show an underlying
  swallowtail, the lower parts of the surface are hidden, so that only
  the point corresponding to the largest complexity is visible for a
  given $(\mu_3, \mu_{16})$ pair. Points A, B, C, and D, marked by
  crosses on a white background, delimit the different branches of the
  putative dynamical G-G transition line (see text for details). }
\end{figure}

Like in the static calculation, the parametric surface
$\{\mu_s(q,m),\mu_p(q,m),\Sigma(q,m)\}$ becomes singular for large
$p-s$, with domains where $\Sigma(\mu_s,\mu_p)$ would be
multivalued. This is illustrated in Fig.~\ref{figdyn} for the $3+16$
model, where one can also see that the surface is more complicated
than in the statics (at this point, it greatly helps to inspect the
surface with an interactive computer graphic software in order to
fully appreciate the nature of the singularities). Following again the
analogy with the van der Waals theory, since the calculation of the
dynamical solution requires that $\Sigma$ should be maximized, a naive
way to define a unique complexity at a given state point
$(\mu_s,\mu_p)$ is to discard the parts of the parametric surface
which do not correspond to the largest possible value of $\Sigma$. In
the case of the $3+16$ model, this reduction leads to the appearance
of three distinct curve portions, successively delimited by points A,
B, C, and D, in Fig.~\ref{figdyn}, which, when crossed, for instance
along lines of constant $\mu_3$, are associated with discontinuities
in $(q,m)$.

Line AB is a portion of the dynamical $m=1$ line
\cite{CriLeuprivate}. Indeed, in the corresponding domain of $\mu_3$
and $\mu_{16}$, the sheet of the parametric surface originating in the
plane $\mu_3=0$ (the right part of the surface in Fig.~\ref{figdyn})
overhangs the one originating in the plane $\mu_{16}=0$ (the left
part). This means in particular that, when point A is approached along
the border between the paramagnetic and glassy phases, the complexity
obtained when coming from the large $q$ side (from the right) is
strictly larger than the one obtained when coming from the small $q$
side (from the left). Thus, when the representative point of the
system crosses the portion of the $m=1$ line comprised between points
A and B, discontinuities in $(q,m)$ (with $m$ jumping to or from unity
depending of the direction of approach) and $\Sigma$ occur, as the
result of jumping from one sheet of the complexity surface to the
other. At point B, the two sheets intersect for the first time. The
$m=1$ line continues below the reduced complexity surface and becomes
irrelevant (at least at the level of the present crude
analysis). Point B marks the beginning of a line of double points,
line BC, along which two distinct $(q,m)$ pairs correspond to the same
value of the complexity of the glassy phase. When this line is
crossed, $(q,m)$ is discontinuous, while $\Sigma$ is
continuous. Eventually, at point C, the line of double points
terminates at a line of cusp singularities which extends towards the
``critical'' point D, providing a final line of $(q,m)$
discontinuities, with a discontinuous $\Sigma$ anew.

As already suggested about the statics, these lines of $(q,m)$
discontinuities appear as candidates for dynamical G-G transition
lines. A simple condition on $s$ and $p$ for their existence is
obtained by looking for stationary points in the dynamical $m=1$ line,
as they are necessary for the presence of a point like point A.  Since
the equations of this $m=1$ line and of the transition line of the
$F_{(s-1)(p-1)}$ schematic model coincide, this condition is just the
same as in the MCT calculation, i.e., $\Lambda(s-1,p-1) > 1$, with
$\Lambda$ defined above.  Thus, candidate dynamical G-G transition
lines are expected in the systems with $s=3$ and $p\ge11$, $s=4$ and
$p\ge16$, or $s=10$ and $p\ge52$, for instance. Generically, keeping
$s$ fixed and increasing $p$, the possibility of a G-G transition line
first appears in the dynamics, then in the statics, so that there is a
whole class of systems like the $3+11$ model which only display a
candidate dynamical G-G transition line. Finally, there are still
limiting cases like the $3+10$ and $10+51$ models, where point A
itself is a stationary point of the dynamical $m=1$ line, to which the
putative dynamical G-G transition line reduces.

So far, we have only considered the topological features of the
complexity surface in the dynamical 1RSB solution and not the actual
values of this function. It came as a surprise in the course of the
present investigation that negative values of the complexity could be
found in a large parameter domain. For a given $s$, this phenomenon is
observed for $p$ larger than the one needed for a singularity to
appear in the complexity surface, so that it does not occur for all
systems with a candidate dynamical G-G transition line. Considering
again the $3+16$ model, it is easily evidenced by computing
$\Sigma(q,m)$ along lines of constant $m$, including $m=1$
\cite{CriLeuprivate}. It results that there is a whole region above
the line joining points C and D in Fig.~\ref{figdyn} and corresponding
to rather small values of $m$, where the only available value of the
1RSB complexity is negative (this is not the case on the $m=1$ line
which falls in the domain where $\Sigma(\mu_s,\mu_p)$ is multivalued
and where, for a given $(\mu_s,\mu_p)$ pair, there is the possibility
of finding a positive value of $\Sigma$ on another sheet of the
complexity surface). As pointed out by Crisanti and Leuzzi
\cite{CriLeuprivate}, this finding signals an unanticipated breakdown
of the 1RSB scheme and calls for a complete reexamination of the phase
behavior of the systems with large $p-s$.

Note that many of the above features of the dynamics of the $s+p$
spin-glass models could have been anticipated from the work of Caiazzo
\textit{et al.}  \cite{CaiConNic04PRL}, who studied the
off-equilibrium dynamics of a lattice gas generalization of the $s+p$
spin-glass model. Indeed, for this richer and more complex system,
these authors found the possibility of dynamical G-G transition lines
and, along these lines, of dynamical singularities marking qualitative
changes in the aging behavior of the system. They also mention that,
deeper in the glassy phase, the ``one-step replica symmetry breaking
solution should not hold and, instead, a spin-glass-like aging
dynamics should be found''. Apparently, and quite surprisingly, the
possibility that the same could be true in the simpler $s+p$
spin-glass model was not envisaged.

In conclusion, guided by previous results on simple MCT models
displaying G-G transitions and higher-order singularities
\cite{FucGotHofLat91JPCM,KraAlb02JCP}, we have proposed a crude
analysis of the 1RSB solution derived by Crisanti and Leuzzi in
Ref.~\onlinecite{CriLeu06PRB} which shows that the phase behavior of
the spherical $s+p$ mean-field spin glasses with well separated $s$
and $p$ does not appear to be fully understood yet. First, there seems
to be a possibility of G-G transition scenarios both in the statics
and the dynamics of these systems. Second, there are clear indications
that the 1RSB solution cannot be correct in the full glassy phase
\cite{CriLeuprivate}. Obviously, it results from the latter finding
that the present analysis based on this 1RSB solution, and also on
probably naive analogies with the van der Waals theory of phase
coexistence, is necessarily at least partly inconsistent. However, it
is sufficient to raise a number of issues about the structure of the
phase diagram of the systems with large $p-s$. It is beyond the scope
of this comment and definitely beyond the expertise of its author to
answer these questions. So, we can only wish that the above
observations will be judged interesting by experts in the theory of
spin glasses and will motivate future studies of these systems, with
potentially interesting new developments, for instance, in the theory
of the amorphous-amorphous transitions in structural glasses.

As a non-expert in spin-glass theory, the author is indebted to
A. Crisanti and L. Leuzzi for useful advices on the presentation of
this work and for insightful observations on the physical implications
of the reported results.


\begin{thebibliography}{99}
\bibitem{lambda} This criterion is obtained by computing the so-called
  exponent parameter $\lambda$ along the transition line. $\lambda$
  should not exceed unity for physically acceptable ideal liquid-glass
  or glass-glass transitions. See Ref.~\onlinecite{LesHouches} for
  instance.
\bibitem{colloids} In fact, it is possible to fully reproduce the MCT
  phenomenology of these complex fluid systems with the present
  schematic models.  For instance, if one keeps $p$ constant and
  considers $q$ as a continuously varying parameter, the sequence of
  phase diagram topologies obtained in the $(v_p,v_q)$ planes by
  increasing $q$, as shown in Fig.~\ref{figmct}, is exactly the same
  as the one obtained in the density-temperature plane of the
  colloidal systems by decreasing the range of the attractive part of
  the interaction.
\end{thebibliography}
\end{document}